\documentstyle[12pt,fleqn]{article}

\author{L.~Didukh and V.~Hankevych \\
{\small \it Ternopil State Technical University, Department of Physics,}\\ 
{\small \it 56 Rus'ka Str., Ternopil UA--46001, Ukraine; 
E-mail: didukh@tu.edu.te.ua}}
\date{}

\date{}
\title{On chemical potential of a generalized Hubbard model with
correlated hopping at half-filling}

\begin{document}
\maketitle
\begin{abstract}
In the present paper we study chemical potential of a generalized Hubbard
model with correlated hopping at half-filling using a generalized
mean-field approximation. For the special case of the model the approach
reproduces the exact results: metal-insulator transition, ground state
energy. Chemical potential of generelized Hubbard model with correlated
hoppig as function of hopping integrals at
zero temperature is found. It is shown that chemical potential of the 
model is temperature-dependent. The dependences of chemical
potential of a generalized Hubbard model with correlated hopping on the
model parameters are different in metallic and insulating phases leading to
a kink at the point of metal-insulator transition in the ground state. 
With the increase of temperature the kink in the chemical potential curve
disappears.The obtained results differ from  those of
the Hubbard model indicating the important role of correlated hopping.
\end{abstract}

\section{Model Hamiltonian}
\setcounter{equation}{0}

One of the simplest model describing correlation effects in narrow energy
bands is the Hubbard model~\cite{1_1}. The 
model Hamiltonian contains two energy parameters: the matrix element
$t_0$ being the hopping integral of an electron from one site to another
($t_0$ is not dependent on occupation of sites involved in the hopping 
process) and the parameter $U$ being the intra-atomic Coulomb repulsion of 
two electrons of the opposite spins. This model is studied intensively (for 
recent reviews see Refs. \cite{1_9,1_10}).

Theoretical analyses, on the one hand, and available experimental data, 
on the other hand, point out the necessity of 
the Hubbard model generalization by taking into account correlated hopping.
This necessity is caused by two reasons. Firstly, theoretical 
analyses~\cite{1_12,1_14} point out the inapplicability of the Hubbard model 
for the description of real strongly correlated electron systems, in some 
compounds (e.g. see the estimation in Refs.~\cite{1_12,1_38}-\cite{1_42}) 
the
matrix element of electron-electron interaction describing correlated 
hopping is the same order that the hopping integral or on-site Coulomb 
repulsion. Secondly, using the concept of correlated hopping and caused by it 
the electron-hole asymmetry we can interpret the peculiarities of some
physical properties of narrow-band materials~\cite{1_12,1_13}-\cite{1_26}.

Now two ways are commonly used to generalize the single-band Hubbard model
by taking into account correlated hopping. One of them has been proposed in
Ref.~\cite{1_14}. Hirsch showed that in contrast to the hopping integral of
the Hubbard model (which is not dependent on occupation of sites involved in
the hopping process) this parameter of a generalized Hubbard model had to
depend on occupation of sites involved in the hopping process.
Hamiltonian of the generalized in such a way Hubbard model is written  as
\begin{eqnarray} \label{G_ham}
&&H=-\sum \limits_{ij\sigma} t_{ij}^{\sigma}a_{i\sigma}^{+}a_{j\sigma}+
U \sum_{i}n_{i\uparrow}n_{i\downarrow},
\\ \label{h_i_G_ham}
&& t_{ij}^{\sigma}=t_{AA}(1-n_{i{\bar \sigma}})(1-n_{j{\bar \sigma}})+
t_{AB}(n_{i{\bar \sigma}}+n_{j{\bar \sigma}}-2n_{i{\bar \sigma}} 
n_{j{\bar \sigma}})+t_{BB}n_{i{\bar \sigma}}n_{j{\bar \sigma}}.
\end{eqnarray}

In recent few years Hamiltonian (\ref{G_ham}) is widely used to study
metal-insulator transition in narrow energy bands \cite{1_16}-\cite{1_20}.

In Ref.~\cite{1_12,1_6} the necessity of the Hubbard model generalization
by taking into account the matrix element of electron-electron interaction
describing intersite hoppings of electrons had been pointed out.
The Hamiltonian of the generalized Hubbard model with correlated hopping is
\begin{eqnarray} \label{ham}
&&H=H_0+H_1+H'_1, \\
&&H_0=-\mu \sum_i \left(X_i^{\uparrow}+X_i^{\downarrow}+2X_i^2\right)+
U\sum_{i}X_i^2,\\
&&H_1=t(n){\sum \limits_{ij\sigma}}'X_i^{\sigma 0} X_j^{0\sigma} +
\tilde{t}(n){\sum \limits_{ij\sigma}}'X_i^{2\sigma}X_j^{\sigma 2},
\\
&&H'_1=t'(n){\sum \limits_{ij\sigma}}' \left(\eta_{\sigma}X_i^{\sigma 0}
X_j^{\bar{\sigma} 2}+h.c.\right),
\end{eqnarray}
where $\mu$ is the chemical potential, $X_i^{kl}=|k\rangle\langle l|$ is
the Hubbard operator~\cite{hubb2}; $|0\rangle$ denotes the state of site, 
which is not occupied by an electron (hole), 
$|\sigma \rangle$ denotes the state of singly occupied (by an electron with 
spin $\sigma$) $i$-site, $|2\rangle$ denotes the state of doubly occupied 
(by two electrons with the opposite spins) $i$-site (doublon).
\begin{eqnarray} \label{c_d_h}
t(n)=t_0+n\sum_{\stackrel{k\neq{i}}{k\neq{j}}}J(ikjk)=t_0+nT_1
\end{eqnarray}
is the effective hopping integral of electrons between nearest-neighbor 
sites of lattice, $n$ is the electron concentration, 
$\eta_{\uparrow}=-1,\ \eta_{\downarrow}=1$,
\begin{eqnarray}
J(ikjk)=\int \int \varphi^*({\bf r}-{\bf R}_i)\varphi({\bf r}-{\bf R}_j)
{e^2\over |{\bf r}-{\bf r'}|}|\varphi({\bf r'}-{\bf R}_k)|^2{\bf drdr'},
\end{eqnarray}
$\varphi({\bf r}-{\bf R}_i)$ is the Wannier function, 
\begin{eqnarray} 
\tilde{t}(n)=t(n)+2T_2, \quad
t'(n)=t(n)+T_2, \quad T_2=J(iiij),
\end{eqnarray}
the prime at the sums signifies that $i\neq j$.

In the model described by Hamiltonian~(\ref{ham}) an electron hopping from 
one site to another is correlated both by the occupation of the sites 
involved 
in the hopping process (with the hopping integral $T_2$) and the occupation 
of the nearest-neighbor sites (with the hopping integral $T_1$) which we 
take into account in the Hartree-Fock approximation (Eq.~(\ref{c_d_h})).
The peculiarity of model~(\ref{ham}) is the concentration dependence 
of the hopping integral $t(n)$ in contrast to similar models.
Below we shall consider the half-filling case, so let us introduce
the following notations: 
$t(n)\equiv t=t_0+T_1,\ \tilde{t}(n)\equiv\tilde{t}=t+2T_2,\ 
t'(n)\equiv t'=t+T_2$.

Properties of generalized Hubbard model with correlated hopping, 
in particular metal-insulator transition (MIT) has been studied in 
a number of recent works~\cite{1_16}-\cite{1_20,1_63}-\cite{1_51}.
At half-filling and $t'=0$ (or $t_{AB}=0$) some exact
results have been found~\cite{1_16,1_63}. In a simple cubic
lattice with coordination number $z$ MIT occurs at
\begin{eqnarray} \label{exact_cr}
U_c=z(|t_{AA}+t_{BB}|)=2z|t_0|.
\end{eqnarray}   
If $U>U_c$ the ground state of system is a paramagnetic Mott-Hubbard 
insulator with the concentration of doubly occupied sites $d=0$, the ground
state energy is equal to zero.

For an arbitrary $t'\ne 0$ (or $t_{AB}\ne 0$) the finding of MIT criterion  
and the description of this phenomenon in a generalized Hubbard model with
correlated hopping still remain an open problem. One of the step to solve
this task is recent papers \cite{1_17,1_20,1_59}-\cite{1_51} where
criteria of MIT, ground state energy, concentration of doubly occupied 
sites have been found. In Refs. \cite{1_17,1_20,1_59}-\cite{1_62} the
authors have obtained the following criterion of MIT:
\begin{eqnarray}
U_c=z(|t_{AA}|+|t_{BB}|)=z(|t|+|\tilde{t}|)
\end{eqnarray}
in agreement with the Mott's general physical ideas~\cite{1_69}.
By means of the slave bosons method~\cite{1_53} it has been found in 
Ref.~\cite{1_51} that MIT occurs at $U_c=4z|t+T_2|$; however here there 
is a problem of
discrepancy of this result with the exact MIT criterion~(\ref{exact_cr}). 

The present paper is devoted to a further study of properties generalized
Hubbard model with correlated hopping at half-filling, in particular
investigation of the model chemical potential in the region of 
metal-insulator transition.   

\section{Results}
\setcounter{equation}{0}

The single-particle Green function in terms of the Hubbard operators reads
as
\begin{eqnarray} \label{spgf}
\langle\langle a_{p\sigma}\!|\hfill a_{p'\sigma}^{+}
\rangle\rangle
=\langle\langle
X_p^{0\sigma}|X_{p'}^{\sigma 0}\rangle\rangle +\eta_{\sigma}\langle\langle
X_p^{0\sigma}|X_{p'}^{2\bar{\sigma}}\rangle\rangle +\eta_{\sigma}\langle\langle 
X_p^{\bar{\sigma} 2}|X_{p'}^{\sigma 0}\rangle\rangle 
%\nonumber\\
+\langle\langle 
X_p^{\bar{\sigma} 2}|X_{p'}^{2\bar{\sigma}}\rangle\rangle.
\end{eqnarray}
The Green function 
$\langle\langle X_p^{0\sigma}|X_{p'}^{\sigma 0}\rangle\rangle$ 
is given by the equation
\begin{eqnarray} \label{eq}
(E+\mu)\langle\langle X_p^{0\sigma}|X_{p'}^{\sigma 0}\rangle\rangle=&&
{\delta_{pp'}\over 2\pi}\langle X_p^{\sigma}+X_p^{0}\rangle+ 
\langle\langle\left[X_p^{0\sigma}, H_1\right]|X_{p'}^{\sigma 0}\rangle\rangle
\nonumber\\
&&+\langle\langle\left[X_p^{0\sigma}, H'_1\right]|X_{p'}^{\sigma 0}\rangle
\rangle,
\end{eqnarray} 
with $[A, B]=AB-BA$. Using a variant~\cite{did} of the generalized
mean-field approximation~\cite{appr} we suppose that
\begin{eqnarray}
\left[X_p^{0\sigma}, H_1\right]=\sum_{j}\epsilon(pj)X_j^{0\sigma},\
\left[X_p^{0\sigma}, H'_1\right]=\sum_{j}\epsilon_1(pj)X_j^{\bar{\sigma}2},
\end{eqnarray}
where $\epsilon(pj)$ and $\epsilon_1(pj)$ are the non-operator expressions. 
The procedure of $\epsilon(pj)$ and $\epsilon_1(pj)$ calculation
is described in Ref.~\cite{did} (here there is a partial equivalence with
the slave boson method~\cite{1_53}).
Thus we obtain the closed system of equations for the Green functions 
$\langle\langle X_{p}^{0\sigma}| X_{p'}^{\sigma 0}\rangle\rangle$ and
$\langle\langle X_{p}^{\bar{\sigma}2}| X_{p'}^{\sigma 0}\rangle\rangle$. 
An analogous procedure is realized also in the equations for the other
Green functions~(\ref{spgf}).

In this way, we find single-particle Green function and quasiparticle
energy spectrum. For paramagnetic case in ${\bf k}$-representation
the spectrum is~\cite{1_61}:
\begin{eqnarray} \label{en_sp}
&&E_{1,2}({\bf k})=-\mu+{(1-2d)(t_{\bf k}+\tilde{t}_{\bf k})+U\over 2}\mp 
{1\over 2}F_{\bf k},
\\
&&F_{\bf k}=\sqrt{\left[B(t_{\bf k}-\tilde{t}_{\bf k})-U\right]^2+
(4dt'_{\bf k})^2},\
B=1-2d+4d^2,
\end{eqnarray}
where $d$ is the doublon concentration which is found from the equation
\begin{eqnarray} \label{d}
d=\langle X_i^2\rangle =&&{1\over N}{\sum_{\bf k}\int\limits_{-\infty}
^{+\infty}}J_{\bf k}(E)dE
\nonumber\\
&&={1\over 2N}\sum_{\bf k}\left(
{A_{\bf k}\over \exp{E_1({\bf k})\over \theta}+1}+
{B_{\bf k}\over \exp{E_2({\bf k})\over \theta}+1}\right),
\end{eqnarray} 
with
\begin{eqnarray}
&&A_{\bf k}={1\over 2}-{B(\tilde{t}_{\bf k}-t_{\bf k})\over 2F_{\bf k}}-
{U\over 2F_{\bf k}}, \\
&&B_{\bf k}={1\over 2}+{B(\tilde{t}_{\bf k}-t_{\bf k})\over 2F_{\bf k}}+
{U\over 2F_{\bf k}},
\end{eqnarray}
$\theta=k_BT$, $k_B$ is the Boltzmann's constant, 
$J_{\bf k}(E)$ is the spectral intensity of the Green function
\begin{eqnarray}
\langle\langle X_p^{\sigma2}|X_{p'}^{2\sigma}\rangle
\rangle_{\bf k}={1\over 4\pi}\left({A_{\bf k}\over E-E_1({\bf k})}+
{B_{\bf k}\over E-E_2(\bf k)}\right).
\end{eqnarray}

For the Hubbard model single-particle Green function~(\ref{spgf}) 
and spectrum~(\ref{en_sp}) at $t_0=0$ have the exact atomic form, 
and at $U=0$ describe the band result. For the case $t'=0$ quasiparticle
energy spectrum~(\ref{en_sp}) reproduces (see Ref.~\cite{1_62}) the 
exact results~\cite{1_16,1_63}: metal-insulator transition and ground state
energy.

Chemical potential of the model is given by the equation 
($\langle X_i^0\rangle=\langle X_i^2\rangle$):
\begin{eqnarray} \label{ch_p}
&&\sum_{\bf k}\left({A_{\bf k}\over \exp{E_1({\bf k})\over \theta}+1}+
{B_{\bf k}\over \exp{E_2({\bf k})\over \theta}+1}\right)
\nonumber\\
&&= \sum_{\bf k}\left({B_{\bf k}\over \exp{-E_1({\bf k})\over \theta}+1}+
{A_{\bf k}\over \exp{-E_2({\bf k})\over \theta}+1}\right).
\end{eqnarray}
From Eq.~(\ref{ch_p}) at $T=0$ we find
chemical potential of generalized Hubbard model with correlated hopping in 
the region of metal-insulator transition as
\begin{eqnarray} \label{ch_p_m}
&&\mu={w\over w+\tilde{w}}U \qquad (U\le w+\tilde{w}), \\
&&\mu ={U\over 2}+{w-\tilde{w}\over 2} \qquad (U>w+\tilde{w}), \label{ch_p_d}
\end{eqnarray}
with  $w=z|t|,\ \tilde{w}=z|\tilde{t}|$.

For the Hubbard model formulae~(\ref{ch_p_m}) and (\ref{ch_p_d}) give
the well-known result $\mu=U/2$. At $t'=0$ chemical potential of the
generalized Hubbard model is equal to $\mu=U/2$
beeing consequence of the electron-hole symmetry which is a characteristic 
of the model in this case. Note that the value of $U_c=w+\tilde{w}$ 
corresponds to the metal-insulator transition point of generalized 
Hubbard model.

At arbitrary value of temperature from Eq.~(\ref{ch_p}) we find the 
expression to calculate chemical potential:
\begin{eqnarray} \label{ch_p_T}
\int\limits_{-w}^{w}\left[{1\over \exp{-E_2(\epsilon)
\over \theta}+1}-{1\over \exp{E_1(\epsilon)\over \theta}+1}
\right]d\epsilon=0,
\end{eqnarray}
where $E_1(\epsilon)$, $E_2(\epsilon)$ is obtained from the respective
formulae~(\ref{en_sp}) for  $E_1({\bf k})$,
$E_2({\bf k})$ as a result of the substitution 
$t_{\bf k}\!\rightarrow\!\epsilon$, 
$\tilde{t}_{\bf k}\!\rightarrow\!{\tilde{t}\over t}\epsilon$, 
$t'_{\bf k}\!\rightarrow\!{t'\over t}\epsilon$.

Fig.~1--3 where the dependences of chemical potential on the ratio $U/w$ and
temperature are plotted show the fact that chemical potential of generalized 
Hubbard model with correlated hopping $\mu >U/2$ and only in the atomic
limit chemical potential $\mu =U/2$. 
From Eqs.~(\ref{ch_p_m}) and (\ref{ch_p_d}) one can see that chemical
potential of generalized Hubbard model with correlated hopping 
depends on $U,\ w,\ \tilde{w}$. The dependences on these paprameters are 
different in metallic and insulating phases, this leads to a kink at the
point of MIT (see Fig.~1); $\tilde{t}=0$ (i.e. $\tilde{w}=0$) and $t'=0.5t$
correspond to the values of correlated hopping parameters
$\tau_1=T_1/|t_0|=0$ and $\tau_2=T_2/|t_0|=0.5$. 

With the increase of temperature the kink in the chemical potential curve
disappears (Fig.~2). The similar peculiarity in the evolution of free energy 
dependence on the parameter $U/w$ with temperature
was noted by Mott~\cite{1_69}. 

From Fig.~3 one can see that in the model under consideration 
chemical potential 
is essentially dependent not only on the parameters $w$ and $\tilde{w}$ and 
also on temperature (in contrast to chemical potential of the Hubbard model), 
and what is more with
the decrease of temperature chemical potential rapidly increases depending on
the correlated hopping paramaters $\tau_1,\ \tau_2$. In 
high temperature region in the generalized Hubbard model chemical
potential tends to $U/2$ with the increase of temperature; really, at 
$T\to \infty$ the probabilities of an electron finding within the lower and
upper Hubbard bands are equal. 

In summary, in the present paper we have studied chemical potential of a
generalized Hubbard model with correlated hopping at half-filling. Chemical
potential of the model as function of the hopping integrals at zero 
temperature has been found. It has been shown that chemical
potential of the model is temperature-dependent. The dependences of chemical
potential of a generalized Hubbard model with correlated hopping on the
model parameters are different in metallic and insulating phases leading to
a kink at the point of metal-insulator transition in the ground state. 
With the increase of temperature the kink in the chemical potential curve
disappears. The obtained result differ essentially from those of the Hubbard 
model that indicate the important role of the correlated hopping.

\subsection{Acknowledgement}
V.H. is grateful to the Max Planck Institute for
the Physics of Complex Systems (Dresden, Germany) for the hospitality 
during the international workshop and seminar on ``Electronic and magnetic
properties of novel transition metal compounds: From cuprates to titanates''
(October 5-31, 1998) where the results considered in the 
present paper were discussed.

Figure~1: The dependence of chemical potential $\mu$ of generalized Hubbard 
model with correlated hopping in the ground state: the upper curve 
corresponds to $\tau_1=0,\ \tau_2=0.5$; the middle curve -- 
$\tau_1=0,\ \tau_2=0.3$; the lower curve -- $\tau_1=\tau_2=0$ (the 
Hubbard model). The asterisks denote the point of metal-insulator 
transition.

Figure~2: Evolution of the dependence of chemical potential $\mu$ of 
generalized Hubbard model ($\tau_1=0,\ \tau_2=0.5$) on $U/w$ with 
temperature $\theta=0,\ 0.1,\ 0.3,\ 1$ respectively. The lowermost curve  
corresponds to the behaviour of chemical potential of the Hubbard model.

Figure~3: The temperature dependence of chemical potential $\mu$ of 
generalized Hubbard model with correlated hopping at $U/2w=1$: the upper
curve corresponds to $\tau_1=\tau_2=0.3$; 
the lower curve -- $\tau_1=\tau_2=0.1$; 
the straight -- $\tau_1=\tau_2=0$ (the Hubbard model).

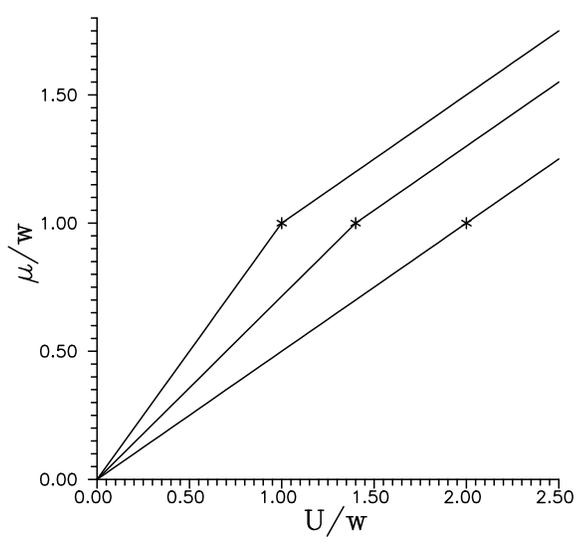
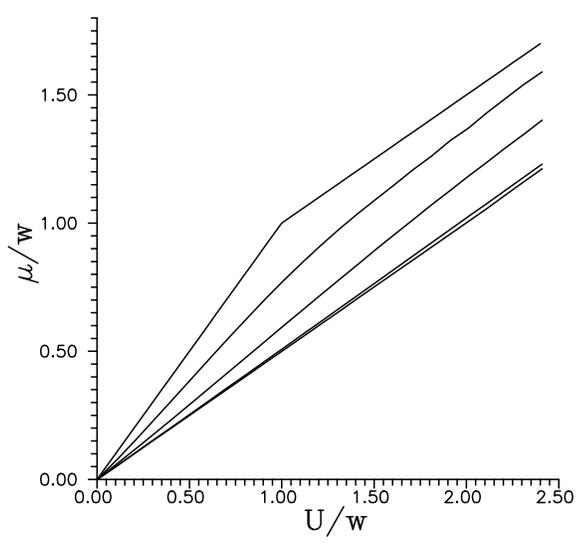
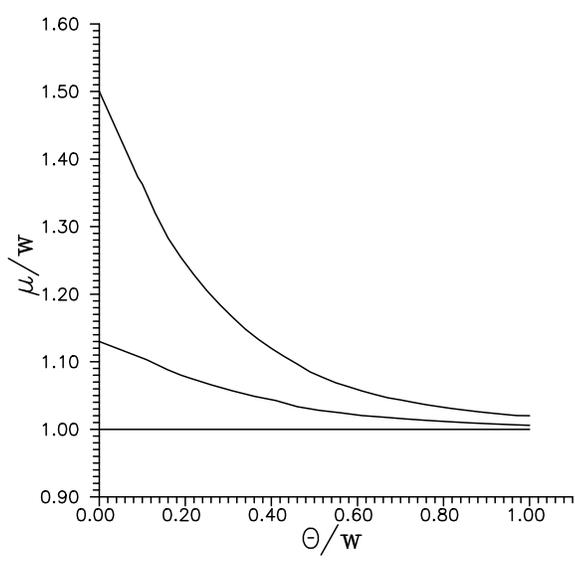
\end{document}